\documentclass[
  DIV18,
  paper=a4,
  fontsize=10pt,
  twocolumn
]
{scrartcl}
\usepackage[utf8]{inputenc}
\usepackage[T1]{fontenc}
\usepackage[english]{babel}
\usepackage{graphicx}
\usepackage{amsmath,amsthm,amsfonts,amssymb}

\usepackage{color}
\usepackage{array}
\usepackage[caption=false]{subfig}
\usepackage{url}
\usepackage{cite}

\usepackage{rotating} %
\usepackage{algorithm}
\usepackage{algpseudocode} %
\algrenewcommand\algorithmicindent{1em}%



\usepackage{authblk}

\usepackage{fancyhdr}

\fancypagestyle{plain}{%
  \fancyhf{}
  \fancyfoot[LE,LO]{\footnotesize Submitted to Computer \& Fluids Special Issue DSFD2017}
  \fancyfoot[CE,CO]{\thepage}
}

\usepackage{listings}
\usepackage[a4paper, colorlinks=true, linkcolor=blue, citecolor=blue, 
urlcolor=blue, bookmarksopen=true, bookmarksnumbered=true]{hyperref}

\newcommand{\img}[1]{images/#1}

\newcommand{\bi}{\begin{itemize}}
\newcommand{\ei}{\end{itemize}}

\newcommand{\be}{\begin{equation}}
\newcommand{\ee}{\end{equation}}

\newcommand{\bibdoiurl}[1]{doi:#1}

\newcommand{\bibarxurl}[1]{arXiv:#1}

\newcommand{\mltwo}[1]{\multicolumn{2}{l|}{#1}}

\begin{document}

\title{Lattice Boltzmann Benchmark Kernels as a Testbed for Performance Analysis}

\author[1]{M. Wittmann}
\author[2]{V. Haag}
\author[1]{T. Zeiser}
\author[2]{H. Köstler}
\author[3]{G. Wellein}

\affil[1]{Erlangen Regional Computing Center, Friedrich-Alexander-University
Erlangen-Nuremberg, Martensstr. 1, 91058 Erlangen, Germany}
\affil[2]{Chair for System Simulation, Friedrich-Alexander-University
Erlangen-Nuremberg, Cauerstr. 11, 91058 Erlangen, Germany}
\affil[3]{Department of Computer Science, Friedrich-Alexander-University
Erlangen-Nuremberg, Matrensstr. 3, 91058 Erlangen, Germany}

\date{October 31, 2017}
\maketitle

\begin{abstract}
Lattice Boltzmann methods (LBM) are an important part of current computational
fluid dynamics (CFD).
They allow easy implementations and boundary handling.
However, competitive time to solution not only depends on the choice of a
reasonable method, but also on an efficient implementation on modern hardware.
Hence, performance optimization has a long history in the lattice Boltzmann
community. 
A variety of options exists regarding the implementation with direct
impact on the solver performance.
Experimenting and evaluating each option often is hard as the kernel itself is
typically embedded in a larger code base.
With our suite of lattice Boltzmann kernels we provide the infrastructure for
such endeavors.
Already included are several kernels ranging from simple to fully optimized
implementations.
Although these kernels are not fully functional CFD solvers, they are equipped
with a solid verification method.
The kernels may act as an reference for performance comparisons and as a blue
print for optimization strategies.
In this paper we give an overview of already available kernels, establish a
 performance model for each kernel, and show a comparison of implementations and
recent architectures.
\end{abstract}

\section{Introduction}

Lattice Boltzmann methods are used in various fields to simulate fluid flows.
Besides different discretizations and collision models a lot of choices arise
concerning data layout, lattice representation, addressing, and propagation step
implementations.
Furthermore, depending on the chosen combination of these options, several
hardware and software optimizations may or may not be possible.
The interaction between all these components can have a severe impact on the
resulting performance.
Experimenting with all these options is often hard as the solvers are
embedded in larger code bases where it is often not possible to easily change
the implementation.
With the \textit{lattice Boltzmann benchmark kernels} we try to create a
framework for an easy evaluation of different data layouts,
addressing schemes, propagation steps, and optimizations.
Furthermore, we deliver several reference implementations of kernels in C which
range from standard unoptimized to highly optimized versions.
The suite is GPLv3 licensed and publicly
available\footnote{\texttt{https://github.com/RRZE-HPC/lbm-benchmark-kernels}}.

The paper is outlined as follows.
In Sect.~\ref{sec:struct} we give a short overview of the LBM
model used, supported architectures, and included sample geometries.
Furthermore, the verification scheme used to validate implementations for their
correctness is presented as well as important points considered regarding
benchmarking.
The available implementations, including the different options regarding data
layout, addressing, etc., are discussed in Sect.~\ref{sec:impl}.
In Sect.~\ref{sec:perf} we establish a performance model and evaluate the
implementations' performance on one selected architecture.
Exemplarily, we discuss several performance effects on this system.
A comparison of the implementation across several current hardware architectures
is presented in Sect.~\ref{sec:pfs}.
Finally Sect.~\ref{sec:conclusion} summarizes the article and gives an outlook
on planned work.

\section{Benchmark Suite}
\label{sec:struct}

Throughout the suite, all benchmark kernels use a D3Q19 discretication with the
two-relaxation-time (TRT) collision model~\cite{ginzburg-2008}.
Depending on the kernel, half or full way bounce back~\cite{haenel-2004} is used.
PDFs are always stored as double precision floating point numbers.

The suite is implemented in C99 and currently focus on the x86-64 platform.
Configurations for recent Linux and GCC/Intel compilers are available.
Some highly optimized kernels use explicit SSE/AVX intrinsics. 
Although we do not include intrinsics for AVX2 or FMA (fused multiply add), we
observed that the Intel compiler transforms AVX intrinsics into these ISA
extensions.

As already noted, the focus of this contribution is clearly on performance
studies and
optimizations.
The suite does not provide a fully-featured flow solver.
This allows us to keep the code modular and easily extensible as we are planning
to provide further kernels as for now only highly optimized ``list''-based
kernels are included (see Sect.~\ref{sec:impl}).

For evaluating the performance with respect to the structure of the simulation
domain, we include several synthetic geometries.
Simple homogeneous arbitrarily scalable geometries like \texttt{channel} or
\texttt{pipe} which represent an empty channel with rectangular or a staircase
approximation of a circular cross section, respectively.
Furthermore, \texttt{blocks} is a heterogeneous geometry
where equidistant blocks of obstacles are distributed over the domain with
adjustable dimensions and distance of the blocks to each other.
The first two geometries can be used to benchmark best-cases as they are
homogeneous and nearly no bounce-back will occur.
The \texttt{blocks} geometry on the other side can be used to study the impact on
performance when the amount of bounce-back is increased and -- depending on the
kernel -- vectorized updates of fluid nodes are no longer completely possible.

\subsection{Verification}
\label{sec:struct:verification}

In order to check if computations performed by a benchmark kernel are correct,
the framework has an inbuilt verification setup. 
For performance reasons this must explicitly be turned on during compilation as
the runtime increases significantly.

For verification a Poision flow is simulated and compared to the analytical
solution.
As in~\cite{pan-2004}, therefore a geometry with periodic boundaries in x and y
direction is setup, i.\,e.\ fluid between two slabs, and on each fluid node a
constant body force is applied. 
After enough iterations, the fluid field in z direction (between the two slabs)
should exhibit a parabola profile and is compared to the analytical solution.

\subsection{Benchmarking}

As benchmarking is a non-trivial endeavor, we try to cover as many problematic
setups as possible already in the implementation:

\paragraph{Thread Affinity}
We support direct pinning (setting the affinity) of threads by the user. 
This binds a thread to the specified core and avoids the thread being migrated
to different cores by the OS.

\paragraph{NUMA placement}
Linux uses by default a first-touch placement strategy. 
This means that a memory page is placed into the NUMA locality domain of the
core first touching it. 
The initialization of the lattice and important accompanying structures respect
this principle by using the same parallelization as the main compute loop later
on, but
correct pinning by the user is required.

\paragraph{Huge Pages}
Current Linux kernels support transparent huge pages. This means that memory
allocated with small $4$\,KiB pages is replaced with larger pages (typically
$2$\,MiB).
In our case, this is in general beneficial as less TLB misses and page walks
occur. 
For this to work, the corresponding OS setting must be enabled to perform this
task every time or on request.
The latter requires an additional call to \texttt{madvise(MADV\_HUGEPAGE)} after
memory allocation.  We perform this for lattice and adjacency list allocation.

\paragraph{Uncovered Points}
Several settings are out of the control of the benchmark kernels themselves, and the user must take care of these,
like fixing the CPU frequency, compiling for the correct architecture,
eventually override the CPU dispatcher, or observe no other tasks are running,
to name just a few.

\section{Implemented Kernels}
\label{sec:impl}

\begin{table*}[t]
\small
\centering
\begin{tabular}{lcccccccc}
\hline

kernel & prop. & data   & addr. & parallel & loop      & padding & \multicolumn{1}{c}{$B_l$} & micro- \\
name   & step  & layout &       &          & blocking  &         & [B/FLUP]  & benchmark \\
\hline
(blk-)push-soa            & OS          & SoA         & D     & x        &  (x)--   & --   & 456       & copy-19       \\
(blk-)push-aos            & OS          & AoS         & D     & x        &  (x)--   & --   & 456       & copy-19       \\
(blk-)pull-soa            & OS          & SoA         & D     & x        &  (x)--   & --   & 456       & copy-19       \\
(blk-)pull-aos            & OS          & AoS         & D     & x        &  (x)--   & --   & 456       & copy-19       \\
aa-soa                    & AA          & SoA         & D     & x        &  x       & --   & 304       & update-19     \\
aa-aos                    & AA          & AoS         & D     & x        &  x       & --   & 304       & update-19     \\
aa-vec-soa                & AA          & SoA         & D     & x        &  x       & --   & 304       & update-19     \\
list-push-soa             & OS          & SoA         & I     & x        &  x       & x    & 528       & copy-19       \\
list-push-aos             & OS          & AoS         & I     & x        &  x       & --   & 528       & copy-19       \\
list-pull-soa             & OS          & SoA         & I     & x        &  x       & x    & 528       & copy-19       \\
list-pull-aos             & OS          & AoS         & I     & x        &  x       & --   & 528       & copy-19       \\
list-pull-split-nt-1s-soa & OS          & SoA         & I     & x        &  x       & x    & 376       & copy-19-nt-sl \\
list-pull-split-nt-2s-soa & OS          & SoA         & I     & x        &  x       & x    & 376       & copy-19-nt-sl \\
list-aa-soa               & AA          & SoA         & I     & x        &  x       & x    & 340       & update-19     \\
list-aa-aos               & AA          & AoS         & I     & x        &  x       & --   & 340       & update-19     \\
list-aa-ria-soa           & AA          & SoA         & I     & x        &  x       & x    & 304--342  & update-19     \\
list-aa-pv-soa            & AA          & SoA         & I     & x        &  x       & x    & 304--342  & update-19     \\
\hline
\end{tabular}
\caption{Implemented kernels with their respective features. Propagation step
OS denotes one step. The addressing
scheme (addr.) is direct (D) for a full array approach or indirect (I) for a list
approach. The loop balance $B_l$ is based on a D3Q19 discretization, double
precision floating point numbers for PDFs and $4$\,b integers for elements of
the adjacency list in case of a list approach.}
\label{tab:impl}
\end{table*}

For implementing a LBM kernel various options regarding lattice representation,
data layout, or propagation step exist. 
In the following we give a short overview of the most common approaches which
we implemented in the benchmark kernels. 
For more details refer
to~\cite{wellein-2006,habich-2012,zeiser-2009-ppl,wittmann-2012-cam,wittmann-2015,wittmann-2016}.
All features of the implemented kernels are summarized in Table~\ref{tab:impl}.

\paragraph{Lattice Representation}
The first option describes how the PDFs of the simulation domain should be
represented.
With the \textit{marker and cell} approach ~\cite{harlow-1965}
a full multi-dimensional array, holding fluid and obstacle nodes, is used together with a
flag field for distinguishing the node types.
PDFs can be accessed by \textit{direct addressing}, i.\,e.\ by simple index
arithmetic.
Especially in porous media like geometries, it is beneficial to store only the
fluid nodes.
These can be arranged in an 1-D vector accompanied by an adjacency
list~\cite{schulz-2001, pan-2004, zeiser-2009-ppl, bernaschi-2008, vidal-2010,
wang-2005, hasert-2013}.
Here, each node's neighbors are accessed via \textit{indirect addressing}, as it
requires a lookup in the adjacency list.
In this work we use the terms ``full array'' and ``list'' for the two
representations, respectively. 
Not covered here are patch-based approaches as used in~\cite{freudiger-2008,
feichtinger-2011} or semi-direct addressing which merges the full array and list
approach as described in~\cite{martys-2002}.

\paragraph{Data Layout}
The data layout can be chosen independently of the lattice representation. 
It describes in which order PDFs of the nodes are contiguously stored in memory.
We concentrate on two incarnations: array-of-structures (AoS) and
structure-of-arrays (SoA).
Sometimes the terms collision optimized layout and propagation optimized layout
are used~\cite{wellein-2006}, respectively.
In the first approach, the PDFs of a node are stored consecutive in memory
whereas in the latter approach PDFs of one direction are consecutive.
Hybrid forms such as AoSoA or SoAoS exist, but are not covered here.

\paragraph{One Step Kernels}
The collision step is defined by the chosen collision model, like
SRT~\cite{chen-1992,qian-1992}, TRT~\cite{ginzburg-2008}, or
MRT~\cite{dhumieres-2002}, but the propagation step provides options for
optimizations.
Typically, collision and propagation are fused into one step and two lattices
are utilized: one source and one destination lattice. 
This kind of implementation is called one step~\cite{pohl-2003} and 
can be implemented in two flavors: push and pull.
With push, the PDFs of a node are read from the source lattice,
collided, and propagated to the neighbor nodes in the destination lattice.
In the case of pull, propagation is performed first, by reading PDFs from the
neighbors in the source lattice, colliding them and then writing them to the
local node in the destination lattice.
One step can be combined with all discussed domain
representations and data layouts.
The corresponding implementations without optimizations are called
\texttt{(list-)push/pull-aos/soa} in the suite.

\paragraph{One Step Kernels with Non-Temporal Stores}
Optimizations regarding one step include avoiding the write allocate for
stores~\cite{wellein-2006}. 
Before a normal store is executed, data residing at the specific location is
first loaded into the cache before being modified. 
This is called write allocate. 
As the written data is not reused until the next time step (and the lattice is
so large that the data in cache will already have been evicted), non-temporal
stores can be used which avoid the extra read operation and thus save memory
bandwidth.
In order to work efficiently, it is best to write in a granularity of complete
cache lines which on all considered architectures are $64$\,b long.
Furthermore, the highest bandwidth with this kind of stores typically allows one
or two concurrent non-temporal store streams only.
We implement this by strip mining the update process to a certain number of
nodes which are loaded, collided, and finally written back to memory with one
or two store streams~\cite{donath-2005,zeiser-2009-ppl}.
We implemented this only for the list representation lattice with SoA data layout
and OS pull, as here stores occur naturally consecutive. 
The kernels implementing these optimizations are
\texttt{list-pull-split-nt-(1s/-s2)-soa}.

\paragraph{AA Pattern Kernels}
The propagation step \texttt{AA pattern}~\cite{bailey-2009} requires only one lattice
and writes only to locations which previously have been read (kernels
\texttt{(list-)aa-soa/-aos}).
It consists of an even and odd time step.
The even time step only requires node local accesses.
Even with a list approach it does not require an indirect access and is
easily (manually) vectorizable with an SoA data layout.
The odd time step reads and writes to neighbors, hence, both require the same
indirect access with a list. 
In an optimized version, we introduce a run length coding where the additional
lookup can be neglected for consecutive nodes, sharing the same access pattern.
This, we call \textit{reduced indirect addressing} or short
RIA~\cite{wittmann-2013-sc13,wittmann-2015}.
As the run length coding describes nodes which can be loaded and stored
contiguously, we use it to implement a partial vectorization (PV) of the odd time
step~\cite{wittmann-2013-sc13,wittmann-2015}.
The latter two optimizations are only implemented for the list version of AA as
\texttt{list-aa-ria-soa} and \texttt{list-aa-pv-soa}.
In the case of a full array implementation, the even and odd time step can be
fully vectorized.
This is done in the \texttt{aa-vec-soa} kernel.

\paragraph{Not Considered Propagation Steps}
Furthermore propagation steps exist, but are (currently) not part of the suite.
An incomplete list includes: i) two-step implementation with split collision and
propagation increase the loop balance dramatically, ii) compressed
grid~\cite{pohl-2003} utilizes only one lattice, traversing it forward and
backwards,  iii) Esoteric Twist~\cite{linxweiler-2011, schoenherr-2011,
pasquali-2016, schoenherr-2015}, which requires only one lattice, exhibits with
indirect addressing the lowest constant loop balance for D3Q19 with
$328$\,B/FLUP, and requires only tree neighbors to be stored in the adjacency
list~\cite{pasquali-2016,schoenherr-2015}, or iv) temporal blocking, were
several updates to a subdomain are performed before it is written back to
memory~\cite{pohl-2003, wellein-2009, habich-2009, nguyen-2010}.

\subsection{Parallelization}

All implemented kernels are OpenMP parallel.
In case of the unblocked one step kernels utilizing a full array (\texttt{push-aos},
\texttt{push-soa}, \texttt{pull-aos}, and \texttt{pull-soa}), the loop nest over the
three spatial dimensions is parallelized with an OpenMP \texttt{parallel for
collapse(3)} clause.
In principal this approach would also support loop blocking, but could lead to
load imbalances. 
In order to avoid this for the blocked variants of these kernels
(\texttt{blk-*}), a different approach is used.
Each thread gets one equally sized part of the domain which is cut in
x direction.
On each subdoamin, loop blocking is then applied.

For the list based kernels, simply the loop over the fluid nodes is parallelized. 
Higher optimized kernels like \texttt{list-pull-split-nt-(1s/-2s)-soa},
\texttt{list-aa-ria-soa}, and \texttt{list-aa-pv-soa} require a manual
scheduling, e.\,g.\ to ensure alignment constraints.

\subsection{Bounce-Back and Periodicity Handling}

Full array kernels iterate over the full domain and initially do not distinguish
between fluid and solid.
In a separate step, bounce-back and correct periodicity of the simulation domain
is treated.
The list kernels allow automatic handling of bounce-back and periodicity via the
adjacency list. 
Hence, they do not need an additional ``correction'' step.

\section{Performance Discussion}
\label{sec:perf}

\begin{table}[tb]
 \small
 \centering
 \begin{tabular}{ll|r}
    \hline
    name      &  & BDW-S     \\
    \hline                                                                                                 
    processor &  & Intel Xeon \\
    name      &  & E5-2630 v4 \\
    \hline
    micro.     &  & Broadwell \\
    \hline
    freq    & [GHz]  & 2.2    \\
    cores   &        & 10     \\
    ISA     &        & AVX2   \\
    sockets &        & 2      \\
    L1 cache & [KiB] & 32     \\
    L2 cache & [KiB] & 256    \\
    L3 cache & [MiB] & 25     \\
    \hline
    \multicolumn{2}{l|}{socket bandwidth} \\
    ~copy          & [GB/s]   & 53.9  \\
    ~copy-19       & [GB/s]   & 48.0  \\ 
    ~copy-19-nt-sl & [GB/s]   & 48.2  \\
    ~update-19     & [GB/s]   & 51.1  \\
    \hline
  \end{tabular}
  \caption{Specifications of the BDW-S system. Bandwidths for \texttt{copy} and
\texttt{copy-19} include write allocate. For \texttt{copy-19-nt-sl} non-temporal
stores are used.}
  \label{tab:hw:meggie}
\end{table}

For discussing the achieved performance of the kernels we firstly introduce the
Roofline performance model.
Afterwards, we present the scaling behaviour of the kernels' performance on 
one socket of the BDW-S system described in Table~\ref{tab:hw:meggie}.
Furthermore, we discuss the impact of loop blocking, padding, and heterogeneous
geometries.
Hereby we limit us to the list based AA pattern kernels.
For all benchmarks in this and the next Sect.\ we fixed the frequency of the
processor to its nominal frequency and set the affinity of each thread
explicitly.
We use Intel C Compiler 17.0.1 and compile with AVX2 and FMA support
enabled.

\begin{figure*}[t]
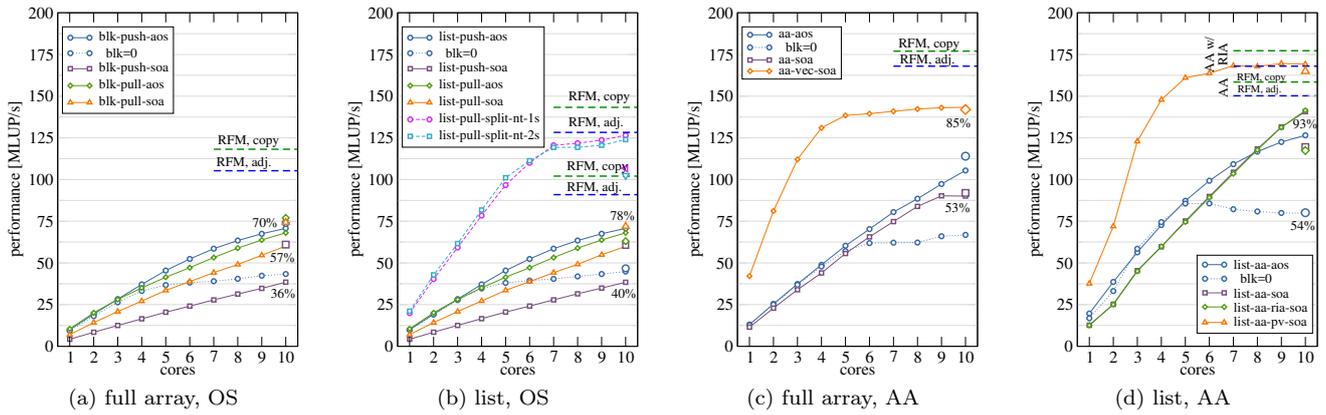
\centering
    \subfloat[full array, OS]{
      \label{fig:p:meggie:array}
      \includegraphics[width=0.22\textwidth,clip=true]{\img{perf-full/perf-meggie}}
    } \, \hfill
    \subfloat[list, OS]{
      \label{fig:p:meggie:list}
      \includegraphics[width=0.22\textwidth,clip=true]{\img{perf-full/perf-meggie-list}}
    } \, \hfill
    \subfloat[full array, AA]{
      \label{fig:p:meggie:array-aa}
      \includegraphics[width=0.22\textwidth,clip=true]{\img{perf-full/perf-meggie-aa}}
    } \, \hfill
    \subfloat[list, AA]{
      \label{fig:p:meggie:list-aa}
      \includegraphics[width=0.22\textwidth,clip=true]{\img{perf-full/perf-meggie-list-aa}}
    } 
    \caption{Performance of kernels on BDW-S system depending on the number of
cores. Single symbols show performance value obtained when SMT was active and
all SMT cores of a socket are used.
Roofline model predictions are given with two different memory bandwidths as
input: (i) the measured copy bandwidth (RFM, copy) and (ii)
the bandwidth measured with the corresponding micro-benchmark for each kernel
from Table~\ref{tab:impl} (RFM, adjusted).
Percentage values show the reached Roofline performance for selected curves. 
    }
    \label{fig:p:meggie}
\end{figure*}

\subsection{Performance Model}
\label{sec:impl:pmodel}

LBM for a D3Q19 discretization, TRT collision model, and double precision
floating point numbers is reckoned to be memory bound.
Therefore, we use the \textit{loop balance} $B_l$ as the relevant metric,
i.\,e.\ the number of bytes that must be transferred between processor and
memory for the update of one fluid lattice node in units of B/FLUP.
We use this as input for the Roofline model~\cite{williams-2009} to derive an
upper performance limit for each kernel.
The loop balance of each kernel is found in Table~\ref{tab:impl} and is taken
from~\cite{wittmann-2012-cam,wittmann-2015}.
Performance $P_\text{max}$ in the memory bandwidth limited case is 
\be
  P_\text{max} = \frac{B}{B_l},
\ee
where $B$ denotes the achievable memory bandwidth.
How much bandwidth is attainable depends heavily on the access pattern.
We use for a general prediction the STREAM \texttt{copy}~\cite{mccalpin-1995}
bandwidth (copy two arrays, without non-temporal stores, including
write-allocate). 
Furthermore, we use three adjusted micro-benchmarks which resemble the nature
of the kernels more specifically: 
\bi
  \item \texttt{copy-19}: concurrently copies 19 arrays (19 read and 19
write streams), 
  \item \texttt{copy-19-nt-sl}: concurrently copies 19 arrays with strip mining
and using only one concurrent non-temporal store stream for write back, and
  \item \texttt{update-19}: updates concurrently 19 arrays.
\ei
Which micro-benchmark is used for which kernel is found in Table~\ref{tab:impl}.
Table~\ref{tab:hw} reports the measured bandwidths of all evaluated systems.

\subsection{Socket Scaling}

Figure~\ref{fig:p:meggie} shows strong scaling of the implemented kernels'
performance depending on the number of cores on one socket of the BDW-S system
from Table~\ref{tab:hw:meggie}.
As geometry \texttt{channel} with dimensions of $500 \times 100 \times 100$
nodes is used.
For using the full socket, the green and blue dashed lines show the Roofline
model predictions for each kernel when the memory bandwidth achieved with
\texttt{copy} or the corresponding micro-benchmark of a kernel (see
Table~\ref{tab:impl}) is used, respectively.
For kernels with AoS data layout performance of several blocking factors is
measured, but only the best performance is reported.

The unoptimized one step kernels in Fig.~\ref{fig:p:meggie:array} with a full array
lattice representation reach a low fraction of the predicted Roofline model
performance, indicated as percentage value for selected kernels in the graph.
Especially \texttt{blk-push-soa} reaches only $36$\,\%. 
By manually padding the array we could lift the performance to $72$\,MFLUP/s.
Measuring the data traffic with likwid~\cite{treibig-likwid-2010} for the bad
case shows a $6$\,\% increased loop balance throughout the memory hierarchy. 
With array padding the measured loop balance between L1/L2 cache is increased by
over $400$\,\% and between L3 cache/memory by about $26$\,\%.
The fraction of L1/L2 TLB misses is in both cases nearly equal. 
This indicates that the initially poor performance was not caused by TLB or
cache thrashing and requires further investigation.
The AoS based kernels require a loop blocking which is discussed in more detail
in the next Sect.~\ref{sec:p:blocking}.
However, for \texttt{blk-push-aos} we exemplarily show the effect. 
The dotted lines in Fig.~\ref{fig:p:meggie}a--d show the performance without
this technique applied, which causes a drop by nearly $60$\,\% in performance.
In general if the lower performance of the kernels would be caused by
under-utilization of core resources, e.\,g.\ introduced through pipeline
bubbles, using SMT threads could help.
Single large symbols indicate the usage of the $20$ virtual cores on BDW-S in
Fig.~\ref{fig:p:meggie}.
Only \texttt{blk-push-soa}'s performance is improved, but below the value we
reached by manual array padding.

The list based kernels in Fig.~\ref{fig:p:meggie:list} show nearly the same
behavior, including the low performance of \texttt{list-push-soa}.
At most $78$\,\% of the maximum attainable performance is reached.
The list based kernels incorporating non-temporal stores
\texttt{list-pull-split-nt-(1s/-2s)-soa} in Fig\.~\ref{fig:p:meggie:list} meet nearly
the prediction.
Using SMT with these kernels shows a contrary effect, as the performance
decreases.
This might be caused by too many non-temporal store streams per physical core.
For the full array kernels we currently do not implement this optimization,
therefore it is not shown.

AA pattern kernels have a significantly lower loop balance compared to the one
step implementations, hence, the Roofline model predictions are higher.
For AA pattern the characteristics between full array and list in
Fig.~\ref{fig:p:meggie:array-aa} and~\ref{fig:p:meggie:list-aa} are different,
respectively.
The full array \texttt{aa-soa} reaches only $53$\,\% of the predicted
performance.
Additional manual padding did not increase the performance.
The \texttt{aa-aos} kernel with loop blocking applied is slightly better and
only \texttt{aa-vec-soa}, the manually vectorized kernel, reaches $85$\,\% of
the Roofline performance.
Measurements show, the latter one saturates the memory bandwidth, but exhibits a
loop balance of $B_l = 323$\,B/FLUP.
Why it is $6$\,\% higher than the theoretical one of $B_l=304$\,B/FLUP is
unclear.
Also this effect requires a more detailed analysis.
The single core performance of \texttt{aa-vec-soa} starts at a much
higher value and saturates early compared to the unoptimized AA pattern kernels.
This effect is caused by the vectorized even and odd time step as the
(theoretical) loop
balance is the same for all AA kernels with full array.
Utilizing SMT is on this machine only beneficial for \texttt{aa-aos}.
The remaining kernels are unaffected. 

List based AA pattern kernels reach a higher performance level.
Here, \texttt{list-aa-soa} gains around $90$\,\% of the prediction and
\texttt{list-aa-aos} is only slightly slower. 
This is interesting as both kernels' loop balance is $B_l=340$\,B/FLUP whereas
the corresponding full array kernels' loop balance is $B_l=304$\,B/FLUP.
The \texttt{list-aa-ria-soa} and \texttt{list-aa-pv-soa} kernels have the same
loop balance which depends on the geometry.
For \texttt{channel} we have $B_l = 305$\,B/FLUP, but
only the latter kernel seems to profit from this optimization, although
measuring the memory traffic indicates both kernels reach the theoretical loop
balance.
The \texttt{list-aa-pv-soa} kernel uses, as \texttt{aa-vec-soa}, 
a vectorized even and partially vectorized odd time step, which causes the
single core performance to start at a higher value and in total saturate with
less cores.
Especially with such an homogeneous domain as \texttt{channel} around $99$\,\%
of the fluid nodes can be updated vectorized in the odd time step.
The usage of SMT has no or even a negative impact on the performance.

Please note that the current implementations of all full array kernels iterate
over all nodes, whether they are fluid or obstacle.
Geometries with a higher fraction of obstacles than \texttt{channel} will cause
a performance drop as only the number of updated fluid nodes is relevant.

\subsection{Loop Blocking} 
\label{sec:p:blocking}

\begin{figure}[tb]\centering
      \includegraphics[width=0.3\textwidth,clip=true]{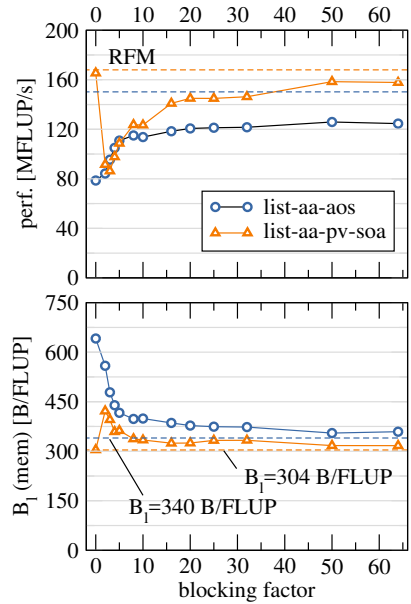}
    \caption{Performance of \texttt{list-aa-aos} and
\texttt{list-aa-pv-soa} on ten cores of BDW-S system depending on blocking
factor. Benchmark geometry is \texttt{channel} with dimensions of $500 \times
100 \times 100$~nodes.  Blocking factor of zero is equal to no blocking
being performed.
    }
    \label{fig:p:blocking}
\end{figure}

It is well known that AoS based kernels require the \textit{layer condition} to
be fulfilled to achieve higher performance~\cite{wellein-2006}.
In the case of \texttt{channel} with dimensions of $500 \times 100 \times
100$ nodes one layer is one y-z-plane of $100 \times 100$ nodes.
During the update of one node in the odd time step of the AA pattern, neighbor
nodes from the next y-z-plane are accessed. 
As we are using an AoS data layout hereby a complete cache line is loaded
containing further PDFs from this node, but only one PDF is used.
The remaining PDFs in the cache line are required, when the corresponding nodes
are reached for update.
To ensure these data are not evicted until needed, a loop blocking is
introduced.
For full array kernels this is achieved by blocking the three loops over the
spatial dimensions.
In case of list kernels the blocking is already performed during setup of the
adjacency list, as the actual update loop iterates over the vector of fluid
nodes.
For \texttt{list-aa-aos} this is only relevant in the
odd time step as only here neighboring nodes are accessed.
For AA pattern at least four layers (including nodes and adjacency list
elements) of each thread must concurrently fit into the (last level) cache.
For BDW-S with $25$\,MiB L3 cache, ten threads, and four layers per thread, a
layer must only contain less than $2900$~nodes or approx.~$54 \times 54$ nodes.
This is also reflected by measurements shown in Fig.~\ref{fig:p:blocking} upper
panel.
With no blocking (blk~$= 0$) performance starts at $80$\,MFLUP/s and
reaches its maximum with blk~$=50$ at around $125$\,MFLUP/s.
With small blocking factors inside a block the layer condition is principally
fulfilled, but they suffer from the high surface to volume ratio.
Updates to nodes at the boundary of a block touch cache lines of the neighboring
blocks in x dimension (also other directions, but x is here the relevant one).
These cache lines are evicted before the corresponding block gets updated.
As the fraction of such cache lines touched at the boundary compared to block
local cache line accesses decreases with an increasing blocking factor
the performance increases as result of an decreasing measured loop balance
$B_l$ (Fig.~\ref{fig:p:blocking} middle panel).

The \texttt{list-aa-pv-soa} kernel with SoA data layouts shows nearly the same
behavior (see~\cite{wittmann-2015} for a detailed analysis).
Also here, just the odd time step is affected as during the even time step only
node local accesses to PDFs occur.
Again the high loop balance with small blocking factors stem from inefficient
cache line usage.
Furthermore, blocking counteracts the fraction of fluid nodes with vectorized
updates during the odd time step.
Without blocking (blk~$=0$) the vectorizability is $v = 98$\,\% and drops to
zero with blocking factor blk~$=2$. 
With increasing blocking factor also $v$ increases again.

\subsection{Array Padding}

\begin{figure}[tb]\centering
  \includegraphics[width=0.45\textwidth,clip=true]{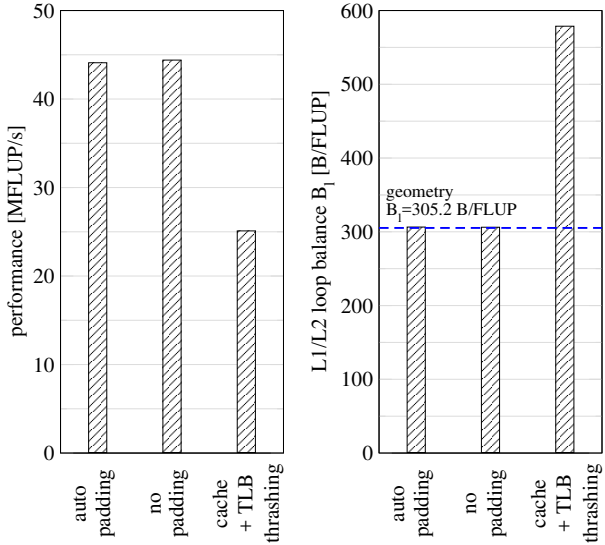}
  \caption{Performance of \texttt{list-aa-pv-aos} on one core of BDW-S system
    with and without padding as well as padding used to provoke cache and (L1) 
    TLB thrashing. 
    Benchmark geometry is \texttt{channel} containing $4\,802\,000$ fluid nodes.
  }
  \label{fig:p:padding}
\end{figure}

Depending on the number of fluid nodes, SoA based kernels can suffer from cache
and/or TLB thrashing which is described for full-array kernels e.\,g.\
in~\cite{wellein-2006} and for list-based kernels in e.\,g.\
in~\cite{wittmann-2016}.
For the SoA data layout relevant is the distance between two PDFs of the same
node. 
In case of the list approach this is the number of fluid nodes. 
If this number maps the cache lines containing the relevant PDFs to only a
subset of cache sets, which number of ways is not enough to hold them all, then
thrashing occurs.
During the process of a node update $19$ ($38$) cache lines for the AA pattern
(OS-based kernels) must be held concurrently in cache.
The BDW-S system's L1 and L2 cache have only eight ways which already can be
critical. 
TLB thrashing is based on the same effect only that the relevant unit is not
cache lines, but pages. Depending on which page size is used, small $4$\,KiB or
huge $2$\,MiB pages, different TLBs might exists.
With padding, we ensure that cache lines are distributed over the sets by
introducing additional nodes which are only used for spacing and are not
updated.

Figure~\ref{fig:p:padding} shows the impact of cache and first level TLB
thrashing on the BDW-S system. 
The \texttt{channel} geometry with $500 \times 100 \times 100$~nodes contains
$4\,802\,000$ fluid nodes which spreads the cache lines and TLB entries over
enough sets, hence no difference between with and without padding is observed.
However, if we pad for cache and TLB thrashing, performance decreases nearly by
$50$\,\% and the measured loop balance between L1 and L2 cache increases far
above the theoretical $B_l = 305.2$\,B/FLUP for this geometry.
Furthermore, first level TLB misses go from practically zero misses/FLUP to
around $2.8$~misses/FLUP.
The number of page-walks caused, i.\,e.\ the actual number of full page table
lookups performed to translate the virtual to physical addresses, does not
differ, as the second level TLB on this architecture is large enough. 
This was different for pre Haswell microarchitectures which only had a first
level TLB for huge pages.

For all list-based kernels by default an automatic padding is activated, where
we try to avoid cache/TLB thrashing.
We assume a TLB cache for $2$\,MiB pages with four sets and a
cache with cache line size $=64$\,B and $512$~sets.
This resembles current Intels first level TLB and L2 cache configurations and
might be counterproductive for other architectures, but can be adjusted by the
user on the command line.

\subsection{Heterogeneous Geometries}

\begin{figure}[htb]\centering
  \includegraphics[width=0.25\textwidth,clip=true]{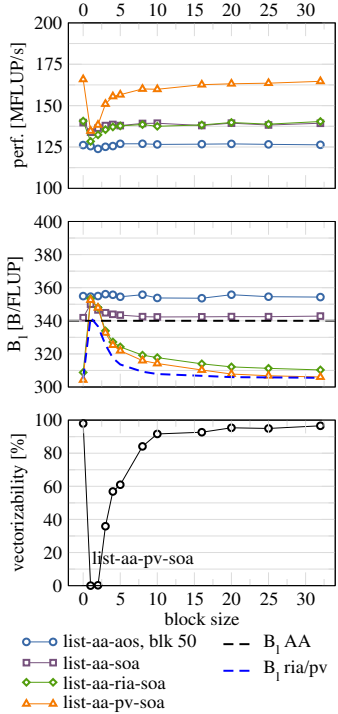}
  \caption{Performance of \texttt{list-aa} kernels with \texttt{blocks} geometry
on ten cores of BDW-S system when the size of blocks is increased.
    Geometry dimensions are $500 \times 100 \times 100$ nodes.
  }
  \label{fig:p:blocks}
\end{figure}

Until now we only used the \texttt{channel} geometry , i.\,e.\ a homogeneous geometry, for
benchmarking where nearly no boundary handling in form of bounce back was
necessary as nearly the whole domain consists of fluid nodes.
Furthermore, this homogeneity causes (i) a low loop balance for
\texttt{list-aa-ria-soa} and \texttt{list-aa-pv-soa} and (ii) guarantees that
most fluid nodes can be updated vectorized in case of \texttt{list-aa-pv-soa}'s
odd time step.

Figure~\ref{fig:p:blocks} shows the performance of the list AA kernels on one
socket of the BDW-S system for the \texttt{block} geometry when the size of the
blocks is increased.
The simple \texttt{list-aa-aos} and \texttt{list-aa-soa} kernels show nearly the
same performance independently of the geometries' structures as their loop
balance is constant.
For small blocks all SoA kernels show a drop in performance for several reasons.
The small blocks have an opposite effect on \texttt{list-aa-ria-soa} and
\texttt{list-aa-pv-soa}, which use the run length coding RIA. 
Instead of having large number of consecutive fluid nodes in an empty channel,
with small blocks the consecutive chunks become very short.
This requires more storage for the run length coding and increases the loop
balance. 
The theoretical values, shown as green dashed line in the middle panel of
Fig.~\ref{fig:p:blocks}, is approximately reached by both implementations
(green and orange lines). 
The performance drop is much more pronounced for \texttt{list-aa-pv-soa}.
Nearly no fluid nodes can be updated vectorized during the odd time step with
small blocks as shown in the lower panel of Fig.~\ref{fig:p:blocks}.

\section{Socket Performance Results on Different Architectures}
\label{sec:pfs}

\begin{table*}[htp]
 \small
 \centering
 \begin{tabular}{ll|rrrrrrrrr}
    \hline
    name      &  & IVB         & HSW-S      & HSW-D     & BDW-S       & SKX        & ZEN-S       & ZEN-D       \\
    \hline                                                                                    
    processor &  & Intel Xeon  & Intel Xeon & Intel Xeon & Intel Xeon & Intel Xeon  & AMD          & AMD \\
    name      &  &  E5-2660 v2 & E5-2695 v3 & E3-1240 v3 & E5-2630 v4 & Gold 6148   & EPYC         & Ryzen 7    \\
              &  &             &            &            &            &             & 7451         & 1700X\\
    \hline
    micro.     &  & Ivy Bridge & Haswell & Haswell & Broadwell        & Skylake     & Zen          &  Zen \\
    \hline
    freq    & [GHz] & 2.2 & 2.3             & 3.4  & 2.2              & 2.4         &  2.3         & 3.4   \\
    cores   &       & 10  & 2 $\times$ 7    & 4    & 10               & 20          &   24         & 8  \\
    ISA     &       & AVX & AVX2            & AVX2 & AVX2             & AVX-512     & AVX2         & AVX2  \\ %
    \mltwo{NUMA LDs}& 1   & 2               & 1    & 1                & 1           & 4            & 1 \\
    \hline                                                        
    L1 & [KiB]      & 32  & 32              & 32   & 32               & 32          & 32           & 32             \\ 
    L2 & [KiB]      & 256 & 256             & 256  & 256              & 1024        & 512          & 512            \\
    L3 & [MiB]      & 25  & 2 $\times$ 17.5 & 8    & 25               & 28          & 8 $\times$ 8 & 2 $\times$ 8  \\
    \hline
    \multicolumn{2}{l|}{socket bandwidth} \\
    ~copy          & [GB/s]  & 39.2 & 52.0 & 22.4 & 53.9 & 102.8 & 130.9 & 29.7  \\
    ~copy-19       & [GB/s]  & 32.7 & 47.3 & 18.8 & 48.0 &  89.7 & 111.9 & 27.2  \\ 
    ~copy-19-nt-sl & [GB/s]  & 35.6 & 47.1 & 19.9 & 48.2 &  92.4 & 111.7 & 27.1  \\
    ~update-19     & [GB/s]  & 37.4 & 44.0 & 19.2 & 51.1 &  93.6 & 109.2 & 26.1  \\
    \hline
    sockets &       & 2   & 2               & 1    & 2                & 2           & 2            & 1 \\
    \hline
  \end{tabular}
  \caption{Specification of systems used for full socket measurements. Number of
cores specifies the physical cores and the number of NUMA locality domains (LD)
are given per socket. The \texttt{copy} and \texttt{copy-19} bandwidths already
accounts for the write allocate. \texttt{copy-19-nt-sl} uses non-temporal stores.
On HSW-S cluster-on-die mode is enabled.
}
  \label{tab:hw}
\end{table*}

\begin{figure*}[htp]\centering
    \subfloat[IVB, 10 cores, AVX]{
      \label{fig:p:fm:emmy}
      \includegraphics[width=0.45\textwidth,clip=true]{\img{perf/perf-full-socket-emmy}}
    } \, \hfill
    \subfloat[HSW-S, 14 cores, AVX2]{
      \label{fig:p:fm:hasep1}
      \includegraphics[width=0.45\textwidth,clip=true]{\img{perf/perf-full-socket-hasep1}}
    } \,
    \subfloat[HSW-D, 4 cores, AVX2]{
      \label{fig:p:fm:woody-hsw}
      \includegraphics[width=0.45\textwidth,clip=true]{\img{perf/perf-full-socket-woody-hsw}}
    } \, \hfill
    \subfloat[BDW-S, 10 cores, AVX2]{
      \label{fig:p:fm:meggie}
      \includegraphics[width=0.45\textwidth,clip=true]{\img{perf/perf-full-socket-meggie}}
    } \,
    \subfloat[SKX, 20 cores, AVX2]{
      \label{fig:p:fm:skylakesp2}
      \includegraphics[width=0.45\textwidth,clip=true]{\img{perf/perf-full-socket-skylakesp2}}
    } \, \hfill
    \subfloat[ZEN-S, 24 cores, AVX]{
      \label{fig:p:fm:naples1}
      \includegraphics[width=0.45\textwidth,clip=true]{\img{perf/perf-full-socket-naples1}}
    } \, 
    \subfloat[ZEN-D, 8 cores, AVX]{
      \label{fig:p:fm:summitridge1}
      \includegraphics[width=0.45\textwidth,clip=true]{\img{perf/perf-full-socket-summitridge1}}
    } \, \hfill
    \subfloat[legend]{%
      \label{fig:p:fm:legend}%
\begin{minipage}[b][5cm][c]{0.45\textwidth}%
\centering%
\small%
\begin{tabular}[b]{rllrl}
\hline
 1 & blk-push-aos && 10 & list-pull-aos   \\
 2 & blk-push-soa && 11 & list-pull-soa   \\
 3 & blk-pull-aos && 12 & list-pull-split-nt-s1 \\ 
 4 & blk-pull-soa && 13 & list-pull-split-nt-s2 \\
 5 & aa-aos       && 14 & list-aa-aos \\
 6 & aa-soa       && 15 & list-aa-soa \\
 7 & aa-vec-soa   && 16 & list-aa-ria-soa \\
 8 & list-push-aos && 17 & list-aa-pv-soa \\
 9 & list-push-soa &&  \\
\hline
\end{tabular}%
\end{minipage}
    } \,
    \caption{Full socket measurements of implemented kernels with
\texttt{channel} geometry of dimensions $500 \times 100 \times100$ nodes on different
hardware architectures. Kernels
supporting blocking several blocking factors are measured, but only the
measurement of the one which resulted in the best performance is reported.
Table~\protect\subref{fig:p:fm:legend} shows the kernels for the corresponding indices
of the plots. Please note the different scaling on the y axis.}
    \label{fig:p:fm}
\end{figure*}

In this section we evaluate the kernels' performance on full sockets of several
hardware architectures specified in Table~\ref{tab:hw}.
Although most systems comprising two sockets, we limit the study to only one.
Despite trying to respect the first-touch policy and thereby being NUMA-aware,
this only works to a certain degree as we will see later on.
Hence, a typical setup for these machines would be using an hybrid OpenMP/MPI
parallel code, where per socket one MPI process is run.
The OpenMP parallel part for such a code, could stem for example from this
suite.

As noted in the previous Sect., we fix the cores' frequency to the nominal
frequency, pin each thread, use Intel C Compiler 17.0.1, and set the target
architecture to compile for to AVX or AVX2, where supported.

In the following we highlight relevant details of the used machines.
BDW-D and ZEN-D resemble ``desktop'' like versions of the processors, which can
nearly saturate the memory bandwidth with one core, but do not show a difference
with utilizing all cores.
For the other machines multiple cores are required to staturate the memory
bandwidth.
All machines except BDW-S and HSW-D have SMT enabled, but we always use the
physical cores only.
Furthermore, HSW-S and ZEN-S exhibit several NUMA locality domains per
socket. 
For these architectures we report also the performance of such a single
domain.
HSW-S is operated in cluster-on-die (CoD) mode, where the processor is logically
divided into two halfs and each half has its own share of the L3 cache and its own
NUMA doamin.
Three cores of ZEN-S share a separate part of the L3 cache. 
Two such complexes, i.\,e.\ six cores build one NUMA domain.
The Skylake SKX system supports AVX-512. However, currently the kernels only
support AVX2, why all reported numbers for this system are measured with AVX2.
We also do not expect any benefit from AVX-512 on this machine as these instructions are
executed with a lower clock frequency and the highly optimized kernels nearly meet the
performance pedictions, i.\,e.\ saturate the memory bandwidth.
On Knights Landing based system, the High Bandwidth Memory (HBM) exhibits a
significantly higher bandwidth than the one achievable from DRAM. Here the
situation might be different. 
It is also the only system where multi-stream memory benchmarks
\texttt{copy-19}, \texttt{copy-19-nt-sl}, and \texttt{update-19} are faster than
the single stream benchmark \texttt{copy}.
The AMD Zen based systems ZEN-S and ZEN-D support AVX2. 
However, measurements on ZEN-S show a slight performance advantage of the
kernels when AVX-only code (without FMA) is used. 
For ZEN-D no difference is measurable.

The kernels' performance on the different machines is found in
Fig.~\ref{fig:p:fm}.
The graphs show in green and blue lines the performance limit
according to the Roofline model from Sec.~\ref{sec:impl:pmodel}.
Hereby for the green lines the \texttt{copy} bandwidth is used, whereas for the blue
lines a micro-benchmark is used which more resembles the kernels access pattern
like \texttt{copy-19}, \texttt{copy-19-nt-sl}, or \texttt{update-19}.
Which bandwidth is used for which kernel is found in Table~\ref{tab:impl}.

As geometry \texttt{channel} with $500 \times 100 \times 100$ nodes is
used.
For kernels supporting blocking several blocking factors are used, but only the
one which results in the best performance is reported.

In general \texttt{list-aa-pv-soa} exhibits the best performance on all
evaluated machines.
Whether the non-vectorized version \texttt{list-aa-ria-soa} with an equal loop
balance reaches the same performance depends on the architectures.
The reason for this is unclear and needs further investigation.
The vectorization in \texttt{list-aa-pv-soa} has actually the effect of
saturating the performance with less cores.
With all cores no difference should be visible as e.\,g.\ HSW-S in
Fig.~\ref{fig:p:fm:hasep1} exhibits.
There is not pattern for the performance behavior of \texttt{list-aa-aos/-soa}.
On HSW-S/-D (Fig.~\ref{fig:p:fm:hasep1} and~\ref{fig:p:fm:woody-hsw}) and
ZEN-S/-D (Fig.~\ref{fig:p:fm:naples1} and~\ref{fig:p:fm:summitridge1}) they
nearly match the performance of their optimized siblings, whereas on the other
machines a larger gap is visible.
The vectorized full array kernel of AA \texttt{aa-vec-soa} reaches despite the
slightly lower loop balance not the performance of \texttt{list-aa-pv-soa}.
Altering the geometry dimensions (not shown), especially increasing the inner
dimension, reveals a significantly higher performance.
As measurements of L2 traffic, branch misspredictions, TLB misses, and performed
page walks for different dimensions show no abnormalities, we speculate that the
relatively short inner dimension of $100$ might be the cause.
The optimization for vector machines as proposed in~\cite{wellein-2006}, where
the loop nest iterating over the three spatial dimensions is fused into one,
could here be beneficial.
The one step kernels with non-temporal stores
\texttt{list-pull-split-nt-(s1/s2)-soa} can be ranked after the optimized AA
pattern kernels.
The measurements show that there is practically no difference between using one
or two non-temporal store streams.
The slowest kernels, as expected by the highest loop balance, are the
unoptimized versions of the one step propagation step.
They show the tendency that the list implementations have a slight advantage.
The measurements however, do not allow for a decision whether SoA or AoS (with
blocking) data layout reaches in general a higher performance.

Typically the bandwidth scales linearly over NUMA locality domains if
first-touch policy is respected.
However, one effect that causes non-perfect scaling for the kernels are pages
which are part of two subdomains that are placed in separate NUMA LDs. 
This is the case when a memory page boundary is not aligned to its subdomain
boundary.
Each core updates a separate subdomain.
The core which first touches this page places it into its NUMA LD.
On HSW-S with CoD mode enabled two NUMA LDs exists. 
All except for one kernel reach between~$90$ and~$99$\,\% of the performance
when we assume the performance of one NUMA LD scales perfectly. 
Only \texttt{blk-push-soa} reaches only $78$\,\%. The reason for this is unclear
and needs further investigation.
On the ZEN-S system with four NUMA domains all kernels reach $90$--$99$\,\% of the
projected performance from one NUMA domain.

\section{Conclusion}
\label{sec:conclusion}

We introduced the \texttt{lattice Boltzmann benchmark kernels}, a suite for
benchmarking simple LBM kernels, which may be used for performance
experiments or can act as a blue print for an implementation.
An initial set of kernels is already implemented, which range from
simple to highly optimized ones.
Thereby widely used data layouts and lattice representations are covered.
The kernels' performance behavior was evaluated by strong scaling on one socket
of a Broadwell based architecture and effects like loop blocking, array padding,
and impact of heterogeneous domains were discussed.
By the usage of the Roofline performance model we were able to determine a
performance limit for each kernel. 
Finally, the kernels were compared on  different current hardware architectures.

We plan to add more optimized kernels, especially with full array as lattice
representation.
These also offer simple ways regarding temporal blocking.
The full array kernels are currently missing a padding mechanism, which will be
implemented, as it is crucial for avoiding cache/TLB thrashing.
Currently the porting of the \texttt{list-aa-pv-soa} kernel to Intel's Knights
Landing architecture is ongoing and will make use of AVX-512 instruction
including gather and scatter operations.

\section*{Acknowledgments}
We would like to thank Christoph Rettinger (Chair for System Simulation), for
helpful discussions regarding verification.
This work was supported by BMBF project SKAMPY (grant no. 01IH15003A) and
KONWHIR (OMI4PAPPS).

\small

\end{document}